\documentclass[
  aps,
  prb,
  showpacs,
  superscriptaddress,
  floatfix,
  twocolumn,
  secnumarabic,
  amssymb,
  amsmath]{revtex4-1}

\usepackage{graphicx}
\usepackage{dcolumn}
\usepackage{bm}
\usepackage{hyperref}

\usepackage{ifthen}
\usepackage{color}
\usepackage{ulem}             
\usepackage{proof}
\newif\ifcom
\newif\ifdel
\comtrue               %

\delfalse

\begin{document}

\title{Temperature Dependent Spin Transport Properties of Platinum Inferred From Spin Hall Magnetoresistance Measurements} 

\author{Sibylle Meyer}
\email{sibylle.meyer@wmi.badw-muenchen.de}
\affiliation{Walther-Mei{\ss}ner-Institut, Bayerische Akademie der Wissenschaften, 85748 Garching, Germany}
\author{Matthias Althammer}
\affiliation{Walther-Mei{\ss}ner-Institut, Bayerische Akademie der Wissenschaften, 85748 Garching, Germany}
\author{Stephan Gepr\"{a}gs}
\affiliation{Walther-Mei{\ss}ner-Institut, Bayerische Akademie der Wissenschaften, 85748 Garching, Germany}
\author{Matthias Opel}
\affiliation{Walther-Mei{\ss}ner-Institut, Bayerische Akademie der Wissenschaften, 85748 Garching, Germany}
\author{Rudolf Gross}
\affiliation{Walther-Mei{\ss}ner-Institut, Bayerische Akademie der Wissenschaften, 85748 Garching, Germany}
\affiliation{Physik-Department, Technische Universit\"{a}t M\"{u}nchen, 85748 Garching, Germany}
\author{Sebastian T. B. Goennenwein}
\affiliation{Walther-Mei{\ss}ner-Institut, Bayerische Akademie der Wissenschaften, 85748 Garching, Germany}
\date{\today}
\begin{abstract}
We study the temperature dependence of the spin Hall magnetoresistance (SMR) in yttrium 
iron garnet/platinum hybrid structures via magnetization orientation dependent magnetoresistance measurements. Our 
experiments show a decrease of the SMR magnitude with decreasing temperature. Using the sensitivity of 
the SMR to the spin transport properties of the normal metal, we interpret our data in terms of a decrease of the spin Hall 
angle in platinum from $0.11$ at room temperature to $0.075$ at $10\,\mathrm{K}$, while the spin diffusion 
length and the spin mixing conductance of the ferrimagnetic insulator/normal metal interface remain 
almost constant.
\end{abstract}
\maketitle
\begin{figure}[tb]
	\includegraphics[width=\columnwidth]{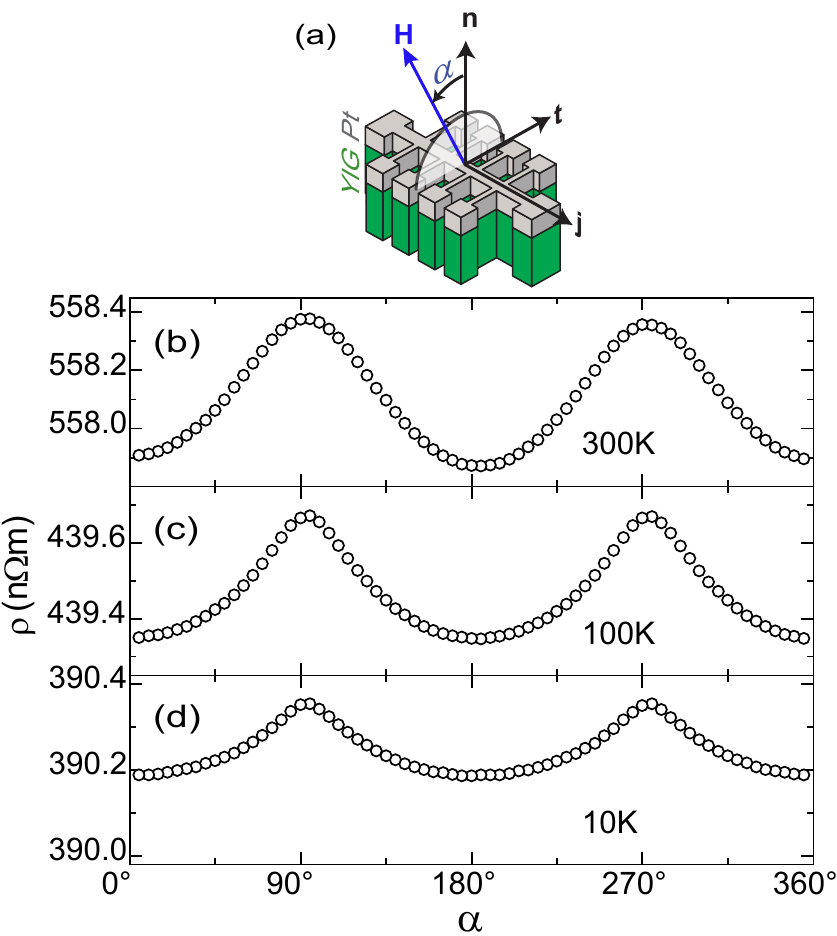}
	\caption{Resistivity $\rho$ versus angle $\alpha$ for a YIG/$3.5\,\mathrm{nm}$ Pt hybrid structure in oopj - ADMR measurements at $1\,\mathrm{T}$, performed at (b) $300\,\mathrm{K}$, (c) $100\,\mathrm{K}$ and (d) $10\,\mathrm{K}$. The different offset values are due to the temperature dependence of the resistivity in the Pt layer [cf.~Fig~\ref{fig:Fig2_rho_SMR_data}(a)]. The sketch in (a) shows the coordinate system defined by $\mathbf {j}$, $\mathbf {t}$, and $\mathbf {n}$ in our YIG/Pt hybrid structures and the definition of the positive rotation angle $\alpha$.}
	\label{fig:Fig1_rho_long_YIG56a_10_300K}
\end{figure}

In a metallic conductor with finite spin-orbit coupling, the flow of electric charge inevitably induces a spin current, and vice versa \cite{Dyakonov_Perel_1971, Hirsch_2004, Kato_Myers_Gossard_Awschalom_2004, Valenzuela_Tinkham_2006}. In the literature, this is usually discussed in terms of the spin Hall effect (SHE), which describes the spin current induced by a charge current, and the inverse spin Hall effect (ISHE),~i.e., the 
charge current arising from a spin current \cite{saitoh_conversion_2006}. The SHE and ISHE are widely exploited for the generation and/or detection of spin currents in ferromagnet/normal metal (FM/NM)
hybrid structures \cite{Bauer_2011},~e.g., in the spin Seebeck effect\cite{uchida_observation_2008, uchida_observation_2010, uchida_spin_2010, jaworski_observation_2010, weiler_local_2012} or in spin pumping experiments\cite{czeschka_scaling_2011, weiler_spin_2011, qiu2012, Tserkovnyak_Brataas_Bauer_2002, Ando_Ieda_Sasage_Takahashi_Maekawa_Saitoh_2009, Costache_Watts_2008, Mosendz_Vlaminck_Pearson_Fradin_Bauer_Bader_Hoffmann_2010}. For a quantitative interpretation of such experiments, a detailed knowledge about the spin transport properties of the respective samples viz. their constituent materials is of key importance. Since any quantitative analysis is complicated by the coexistence of electronic and magnonic spin currents, hybrid devices based on ferromagnetic insulators (FMI) came into focus, and resulted in particular in a renewed interest in the ferrimagnetic insulator yttrium iron garnet ($\mathrm{Y}_3\mathrm{Fe}_5\mathrm{O}_{12}$, YIG)\cite{uchida_observation_2010, uchida_spin_2010, weiler_local_2012,huang_transport_2012, Nakayama2013, VlietstraSMR2013, Alti,Lu_MR_2013, Isasa2013}. The characteristic magnetoresistive effect reported from YIG/Pt (FMI/NM)
heterostructures by different groups\cite{weiler_local_2012, huang_transport_2012, Nakayama2013, VlietstraSMR2013, Alti, Lu_XMCD_2013, Lu_MR_2013, Isasa2013}, however, is controversially discussed. Huang \textit{et al}.\cite{huang_transport_2012} and Lu \textit{et al}.\cite{Lu_XMCD_2013, Lu_MR_2013} ascribe the observed magnetoresistance to a static magnetic proximity effect in Pt. On the other hand, the magnetoresistance in FMI/NM hybrids 
can also be understood as a spin current-based effect, the so-called spin Hall magnetoresistance (SMR) \cite{Nakayama2013,Alti,ChenSMR2013}. This interpretation naturally accounts for both the magnetization orientation dependence and the magnitude of the observed magnetoresistance\cite{weiler_local_2012,Nakayama2013,Alti,ChenSMR2013, Gepraegs_YIG2012}.\\
In this letter, we experimentally study the temperature-dependent evolution of the magnetoresistance in a set of YIG/Pt bilayer samples with different Pt thicknesses, and interpret our observations in terms of the SMR. We extract the magnitude of the SMR effect from magnetoresistance measurements as a function of the magnetization orientation (angle dependent magnetoresistance, ADMR). The ADMR data recorded in the temperature range ${10}\,\mathrm{K}\le T \le 300\,\mathrm{K}$ consistently show that the SMR magnitude decreases with decreasing temperature. Using the SMR theory\,\cite{ChenSMR2013}, we extract the effective spin diffusion length $\lambda(T)$ in Pt, the real part of the spin mixing conductance $G_{\mathrm{r}}(T)$ of the YIG/Pt interface, as well as the spin Hall angle $\theta_{\mathrm{SH}}(T)$ in Pt. We find that $\lambda$ and $G_{\mathrm{r}}$ are about independent of temperature, while  $\theta_{\mathrm{SH}}$ decreases from $\theta_{\mathrm{SH}}\approx 0.11$ at $300\,\mathrm{K}$ to $\theta_{\mathrm{SH}}\approx 0.075$ at $10\,\mathrm{K}$.\\
The SMR arises from the absorption ($\mathbf {M}\perp\mathbf {\sigma}$) or reflection ($\mathbf {M}\parallel\mathbf {\sigma}$) of a spin current at the FMI/NM interface and thus depends on the orientation of the magnetization 
$\mathbf {M}$ of the FMI with respect to the spin polarization $\sigma$ of the spin current\cite{ChenSMR2013}. This results in a characteristic dependence of the resistivity $\rho$ of the NM layer on the orientation $\mathbf{m}=\mathbf{M}/\left|\mathbf{M}\right|$ of the magnetization in the adjacent FMI:\cite{weiler_local_2012,Nakayama2013,Alti,VlietstraSMR2013,ChenSMR2013}
\begin{equation}
\rho=\rho_0 + \Delta \rho \left(\mathbf{m} \cdot \mathbf{t} \right)^2=\rho_0 + \Delta \rho \sin^2 \alpha,
\label{eq:rho}
\end{equation}
with the SMR amplitude
\begin{equation}
\frac{\Delta \rho}{\rho_0}=-\frac{2\theta_{\mathrm{SH}}^2\lambda^2\frac{\rho}{t}G_{r}\tanh^2\left(\frac{t}{2\lambda}\right)}{1+2\lambda\rho G_{r}\coth\left(\frac{t}{\lambda}\right)}.
\label{SMR_Nakayama}
\end{equation}
Here, $\rho_0$ is the  intrinsic electric resistivity of the NM layer and $\Delta \rho$ is the magnitude of the magnetization-orientation dependent resistivity change arising from the interplay of charge and spin currents at the FMI/NM interface, $\mathbf{t}$ is a unit vector orthogonal to both the direction $\mathbf{j}$ of charge current flow and the film normal $\mathbf{n}$ (see Fig.~\ref{fig:Fig1_rho_long_YIG56a_10_300K}), and $\alpha$ is the angle enclosed by $\mathbf{n}$ and the magnetization orientation $\mathbf{m}$. 
As evident from Eq.\,(\ref{SMR_Nakayama}), the SMR varies characteristically with the thickness $t$ of NM \cite{Nakayama2013,huang_transport_2012, Alti, VlietstraSMR2013}. Thus, the measurement of the SMR as a function of $t$ allows for a quantitative evaluation of both $\theta_{\mathrm{SH}}$ and $\lambda$ of the NM. Since we here study Pt films with thicknesses down to $1\,\mathrm{nm}$, we explicitly take surface scattering effects into account by considering that the resistivity $\rho = \rho(t)$ depends on the Pt film thickness\cite{Alti}.\\
The samples used in our experiments are YIG/Pt thin film heterostructures deposited onto (111)-oriented gadolinium gallium garnet (GGG) or yttrium aluminum garnet (YAG) single crystal substrates as described earlier\cite{Gepraegs_YIG2012}. The YIG thin films with a thickness of approximately $ 60\,\mathrm{nm}$  were epitaxially grown via pulsed laser deposition from a stoichiometric polycrystalline target, utilizing a KrF excimer laser with a wavelength of $248\,\mathrm{nm}$ at a repetition rate of $10\,\mathrm{Hz}$. 
The deposition was carried out in an oxygen atmosphere at a pressure of $25\times 10^{-3}\,\mathrm{mbar}$ and a substrate temperature of $500^{\circ}\mathrm{C}$ (YAG) or $550^{\circ}\mathrm{C}$ (GGG), respectively. 
After cooling the sample to room temperature, we in-situ deposited a polycrystalline Pt layer of thickness $t$ via electron beam evaporation on top of the YIG film. We applied high-resolution X-ray reflectometry (HR-XRR) to determine $t$ for all samples, using the Software Package LEPTOS (Bruker AXS), (see~Tab.~\ref{tab:samples}).
We patterned the YIG/Pt bilayers into Hall bar structures (width $w=80\,\mathrm{\mu m}$, contact separation $l=600\,\mathrm{\mu m}$) using optical lithography and argon ion beam milling (see Fig.~\ref{fig:Fig1_rho_long_YIG56a_10_300K}(a)), and mounted them in the variable temperature inset of a superconducting magnet cryostat for magnetoresistance measurements ($10\,\mathrm{K}\leq T \leq 300\,\mathrm{K}$). We performed ADMR measurements\cite{limmer_angle-dependent_2006, limmer_advanced_2008, Alti} by rotating an external magnetic field of constant magnitude $\mu_0 H\leq 7\,\mathrm{T}$ in the plane perpendicular to the current direction $\mathbf j$ (oopj geometry) and recording the evolution of the sample's resistivity $\rho\left( \alpha_{\mathrm{H}} \right)$. Here, $\alpha_{\mathrm{H}}$ denotes the angle between the magnetic field $\mathbf{H}$ and the surface normal $\mathbf{n}$. The magnitude of the magnetic field is intentionally chosen much larger than the anisotropy and demagnetizing fields of YIG, in order to ensure that the YIG magnetization $\mathbf{M}$ is always saturated and oriented along $\mathbf{H}$, $\alpha_{\mathrm{H}} = \alpha$. 
By choosing the oopj rotation geometry, we can separate the SMR signal from an anisotropic magnetoresistance (AMR) in the polycrystalline Pt layer. In particular, one would not expect a magnetization orientation dependence of the resistivity in this configuration for AMR, as discussed in \cite {Lu_MR_2013, Alti}”.\\
The longitudinal resistivity $\rho\left( \alpha \right)=V_{\mathrm{long}}\left( \alpha \right) /(J l)$ of the sample can then straightforwardly be calculated from the voltage drop $V_{\mathrm{long}}\left( \alpha \right)$ along the direction of charge current flow and the magnitude $J$ of the charge current density.
\begin{table}[tb]
\begin{center}
		\begin{tabular}{c c c c c c}
  \hline
	\hline
  substrate & $t (\mathrm{nm})$ & $h (\mathrm{nm})$ & substrate & $t (\mathrm{nm})$ & $h (\mathrm{nm})$\\
			\hline 
  YAG & 0.8 & 0.7 & GGG & 3.5 & 0.7\\
  YAG & 2.0 & 0.8 & YAG & 6.5 & 0.9 \\
  GGG & 2.2 & 0.7 & GGG & 11.1 & 0.6\\
  GGG & 2.5 & 0.5 & GGG & 17.2 & 0.6\\
  YAG & 3.0 & 0.8 & YAG & 19.5 & 1.0\\
	\hline
	\hline
		\end{tabular}
	\caption{Substrate material, platinum thickness $t$ and interface roughness $h$ (rms value) for all YIG/Pt bilayer heterostructures investigated in this work.}
	\label{tab:samples}
\end{center}
\end{table}
Figure~\ref{fig:Fig1_rho_long_YIG56a_10_300K} shows a typical set of $\rho\left( \alpha \right)$ ADMR curves, recorded in the YIG/Pt sample with $t=3.5\,\mathrm{nm}$ at different, constant temperatures while rotating a magnetic field $\left| \mu_{0} \mathbf{H} \right|=1\,\mathrm{T}$. In a series of ADMR measurements at different magnetic fields (not shown here), we furthermore checked that $\rho\left( \alpha \right)$ does not depend on the field magnitude for $0.5\,\mathrm{T}\le \mu_{0}H \le 7\,\mathrm{T}$. As evident from Fig.~\ref{fig:Fig1_rho_long_YIG56a_10_300K}, the measured resistivity shows a $\sin^2(\alpha)$-behavior with respect to the magnetization orientation, as also reported in earlier SMR experiments\cite{Alti}.
\begin{figure}[b]
\includegraphics[width=\columnwidth]{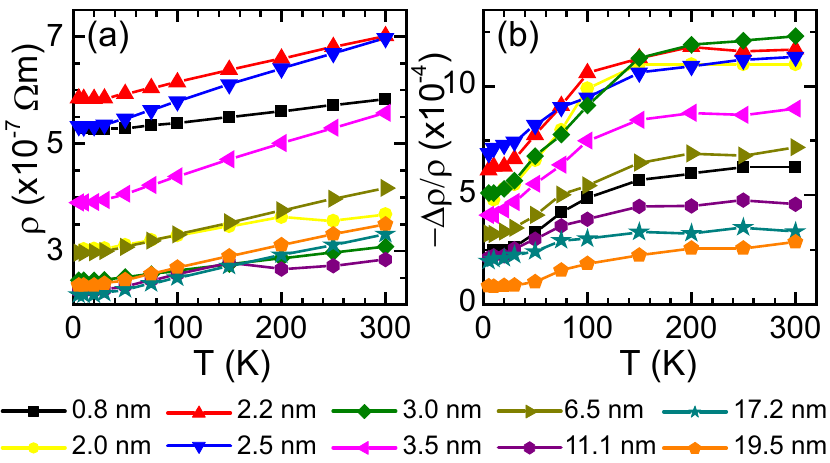}
	\caption{Temperature dependence of (a) the resistivity $\rho$ and (b) the SMR signal $-\Delta \rho/\rho_0$ in YIG/Pt with different values $t$  of the Pt thickness at $\mu_0\mathrm{H}=1\,\mathrm{T}$. The lines are guides to the eye.}
	\label{fig:Fig2_rho_SMR_data}
\end{figure}
We now address $\rho$ of the normal metal Pt in more detail. We observe an increase of $\rho$ with decreasing $t$, which we attribute to the finite roughness of the YIG/Pt interface. Upon decreasing the temperature from room temperature to $10\,\mathrm{K}$, $\rho$ decreases by a factor of about $1.5$
[cf.~Fig.~\ref{fig:Fig2_rho_SMR_data}(a)] for all samples as expected for metals. In order to take the film thickness and temperature dependence of $\rho$ into account, we use a thickness dependent resistivity~\cite{VlietstraSMR2013, Castel2012} $\rho(t, T)$\cite{Fischer1980}:
\begin{figure}[tbh]
	\centering
		\includegraphics[width=\columnwidth]{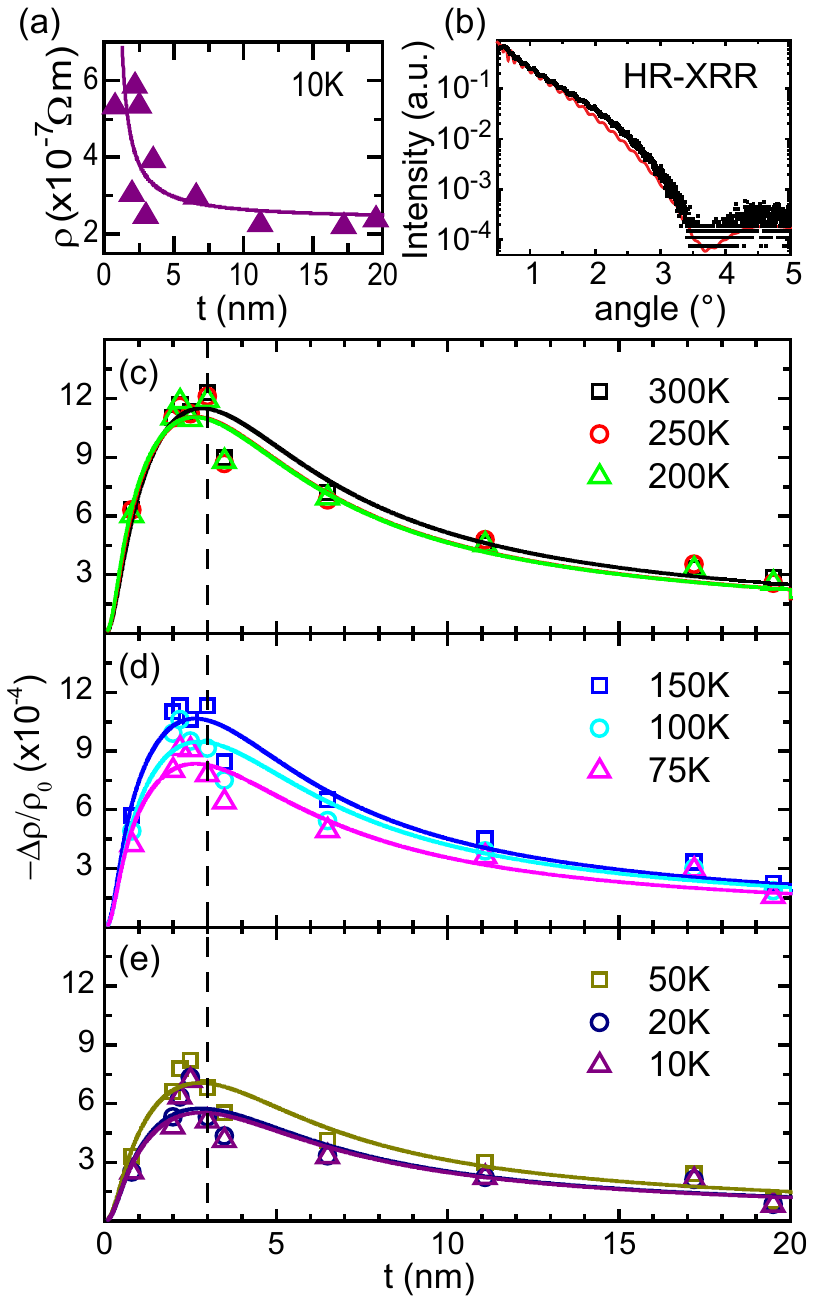}
	\caption{ (a) Thickness dependence of $\rho$ (symbols) for the $10\,\mathrm{K}$ data and a fit using Eq.\,(\ref{eq:rho_T}) (solid line). Panel (b) shows exemplary HR X-ray reflectometry (HR-XRR) data for a sample with a Pt thickness of $t = 2.5\,\mathrm{nm}$. The red line shows a HR-XRR fit for a thickness of the YIG layer of $t_{\mathrm{YIG}} = 53.4\,\mathrm{nm}$ and a roughness of $h = 0.49\,\mathrm{nm}$. The SMR effect $-\Delta\rho/\rho_0$ in YIG/Pt bilayers plotted versus the Pt thickness $t$ at temperatures between $300\,\mathrm{K}$ and $10\,\mathrm{K}$ is shown in panels (c)-(e). The symbols represent the experimental data taken at $\mu_0 H=1\,\mathrm{T}$ and the solid lines depict the SMR calculated from Eq.~(\ref{SMR_Nakayama}) using the parameters $\theta_{\mathrm{SH}}$, $G_{\mathrm{r}}$ and $\lambda$ shown in Fig.~\ref{fig:Fig4_Theta_lambda_simu_res}. The black dashed line indicates the maximum of the observed SMR at $t\approx3\,\mathrm{nm}=2\lambda$.}
	\label{fig:Fig3_SMR_data_vs_Simu}
\end{figure}
\begin{equation}
\rho(t,T) =\rho_{\infty}(T)\left(1+\frac{3}{8\left(t-h\right)} \ell_{\infty}\left(1-p\right)\right) ,
\label{eq:rho_T}
\end{equation}
where $\rho_{\infty}$ is the resistivity for $t\rightarrow\infty$, $h$ the rms interface roughness, $\ell_{\infty}$ the mean free path for $t\rightarrow\infty$ and $p$ the fraction of electrons scattered at the metal surface. Here we assume a diffusive limit ($p=0$) and choose $\rho_{\infty}(T) =\rho(19.5\,\mathrm{nm}, T)$ (the thickest film studied is assumed to be bulk-like) and $\ell_{\infty}=3\,\mathrm{nm}$ from a fit of Eq.~(\ref{eq:rho_T}) to the experimental data as exemplarily shown in Fig.\,\ref{fig:Fig3_SMR_data_vs_Simu}(a) for the $10\,\mathrm{K}$ data. To enable a straightforward fit of the data as a function of the film thickness, i.e., across several samples, we use one and the same average rms value of $h=0.7\,\mathrm{nm}$ for the interface roughness (derived from HR-XRR as listed in tab.\,\ref{tab:samples}) for all samples.
As evident from Fig.~\ref{fig:Fig2_rho_SMR_data}(b), the magnitude of the SMR signal $\Delta \rho / \rho_0$ decreases with decreasing temperature for all samples. Upon plotting $\Delta \rho/\rho_0$ as a function of $t$ for different $T$ as shown in Fig.~\ref{fig:Fig3_SMR_data_vs_Simu}(c)-(e), a clear maximum in the SMR signal magnitude at around $t \approx 3\,\mathrm{nm}$ becomes evident. Note that according to Eq.~(\ref{SMR_Nakayama}) the SMR should show a maximum at $t \approx 2\lambda$. Fig.~\ref{fig:Fig3_SMR_data_vs_Simu}(c)-(e) reveals that this maximum appears at the same $t$ value of about $3\,\mathrm{nm}$ for all temperatures within the accuracy of our measurements, suggesting that the spin diffusion length $\lambda$ is only weakly temperature dependent.
\begin{figure}[tbh]
	\centering
		\includegraphics[width=\columnwidth]{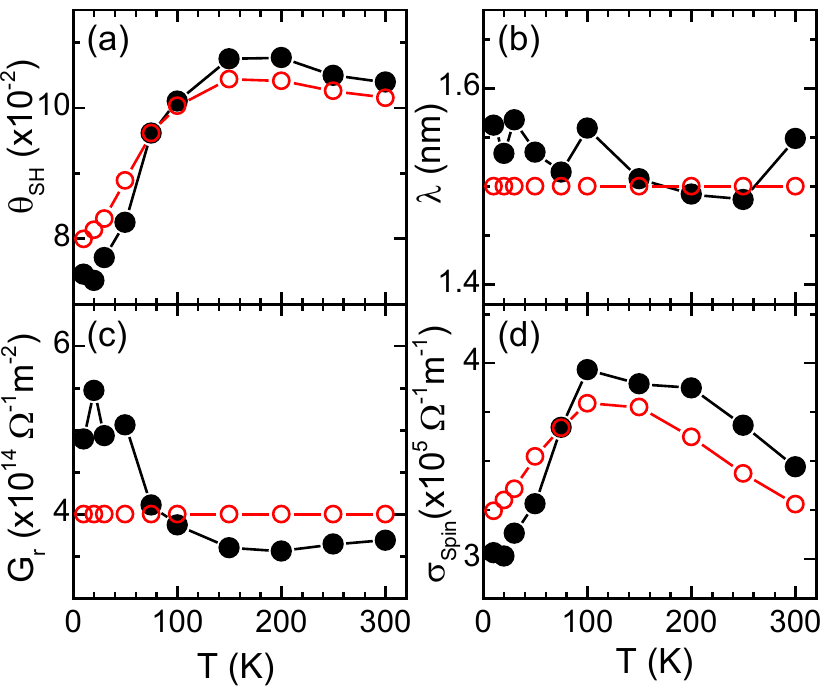}
	\caption{Temperature dependence of (a) the spin Hall angle $\theta_{\mathrm{SH}}$, (b) the spin diffusion length $\lambda$ and (c) the spin mixing conductance $G_{\mathrm{r}}$ for Pt extracted from a fit to our SMR data. Full black symbols represent the values obtained using three free parameters $\theta_{\mathrm{SH}}(T)$, $G_{\mathrm{r}}(T)$ and $\lambda(T)$, red open symbols indicate simulations with constant $\bar G_{\mathrm{r}} = 4 \times 10^{14}\,\Omega^{-1} \mathrm{m}^{-2}$ and $\bar \lambda = 1.5\,\mathrm{nm}$. Panel (d) shows $\sigma_{\mathrm {spin}}$ calculated using the temperature dependent resistivity $\rho(t)$ from our experimental data for a sample with $t=3\,\mathrm{nm}$.}
	\label{fig:Fig4_Theta_lambda_simu_res}
\end{figure}
Finally, we use Eq.~(\ref{SMR_Nakayama}) to extract the Pt spin transport parameters from our set of experimental data. As discussed above, Eq.~(\ref{SMR_Nakayama}) depends on four parameters: $\theta_{\mathrm{SH}}(T)$, $\lambda(T)$, $\rho(t, T)$ and $G_{\mathrm{r}}(T)$. Since we use $\rho(t, T)$ calculated from Eq.~(\ref{eq:rho_T}), this leaves $\theta_{\mathrm{SH}}(T)$, $\lambda (T)$, and $G_{\mathrm{r}}(T)$ as free parameters. Fitting the data then yields $\theta_{\mathrm{SH}}(T)$, $\lambda (T)$ and $G_{\mathrm{r}}(T)$ as given by the full symbols in Fig.~\ref{fig:Fig4_Theta_lambda_simu_res}. The parameters consistently describe our entire set of experimental data, as depicted by the solid lines in Fig.~\ref{fig:Fig3_SMR_data_vs_Simu}(c)-(e). As the temperature dependence of $G_{\mathrm{r}}$ and $\lambda$ is rather weak and comparable to the fitting error, we performed a second analysis with temperature independent $\bar G_{\mathrm{r}} = 4.0\times10^{14}\,\Omega^{-1} \mathrm{m}^{-2}$ and $\bar{\lambda} = 1.5\,\mathrm{nm}$ values [cf.~Fig.~\ref{fig:Fig4_Theta_lambda_simu_res}(c)]. The $\theta_{\mathrm{SH}}(T)$ values obtained from this simple analysis [cf.~red open symbols in Fig.~\ref{fig:Fig4_Theta_lambda_simu_res}(a)] are very similar to the ones obtained from the full fit. This suggests that the real part of the spin mixing conductance $\bar G_{\mathrm{r}}$ is almost independent of temperature, as one might naively expect considering that the density of states in Pt does not significantly change with $T$. The spin diffusion length $\bar{\lambda}$ obtained from our fit is comparable to earlier results\cite{Ulrichs2013}. However, since the spin diffusion strongly depends on the density and type of impurities in the NM, a significant difference of values for $\lambda$ spreading from $1.25\,\mathrm{nm}$ \cite{Ulrichs2013} to $(14 \pm 6)\,\mathrm{nm}$ \cite{Kurt2002} is reported in the literature.\\
From the relation $\theta_{\mathrm{SH}}=\sigma_{\mathrm{spin}}/\sigma$, we can calculate the temperature dependent spin Hall conductivity $\sigma_{\mathrm{spin}}(T)$ using the temperature dependent $\theta_{\mathrm{SH}}(T)$ from the simulation and the measured electrical conductivity $\sigma (t, T)=\rho^{-1}(t, T)$. Figure~\ref{fig:Fig4_Theta_lambda_simu_res}(d) shows $\sigma_{\mathrm{spin}}(T)$ exemplary for the $t=3\,\mathrm{nm}$ sample [the $\rho(T)$-evolution is very similar in all samples studied, see Fig.~\ref{fig:Fig2_rho_SMR_data}(a)]. From both simulation approaches, we obtain a $\sigma_{\mathrm{spin}}(T)$ dependence that does not substantially change within the temperature range investigated, with a magnitude $\sigma_{\mathrm {spin}}=(3.6\pm 0.3)\times 10^5\,\Omega^{-1}\mathrm{m}^{-1}$ quantitatively consistent with other measurements~\cite{Kimura_Otani_Sato_Takahashi_Maekawa_2007}.\\
In summary, we have investigated the SMR in YIG/Pt heterostructures with different Pt thicknesses via ADMR measurements at temperatures between $10\,\mathrm{K}$ and room temperature. We observe a decrease of the SMR at low temperatures for all Pt thicknesses. We used the SMR theory to extract the temperature dependence of the spin mixing conductance $G_{\mathrm{r}}$ for the YIG/Pt interface, as well as the spin Hall angle $\theta_{\mathrm{SH}}$ and the spin diffusion length $\lambda$ in Pt. Our data suggests $\lambda$ and $G_{\mathrm{r}}$ to be almost $T$-independent, while $\theta_{\mathrm{SH}}$ decreases from $0.11$ at room temperature to $0.075$ at $10\,\mathrm{K}$. Nevertheless, the spin Hall conductivity in Pt does not substantially change as a function of temperature, with $\sigma_{\mathrm{spin}}=(3.6\pm 0.3)\times 10^5\,\Omega^{-1}\mathrm{m}^{-1}$ .\\

We thank T. Brenninger for technical support, A. Erb for the fabrication of the stoichiometric YIG target and G.E.W. Bauer and M. Schreier for fruitful discussions.
Financial support by the Deutsche Forschungsgemeinschaft via SPP 1538 (project no. GO 944/4) and the German Excellence Initiative via the "Nanosystems Initiative Munich (NIM)" is gratefully acknowledged.

\end{document}